\documentclass[a4paper,11pt]{article}
\usepackage{pos}
\usepackage{xfrac}
\usepackage{braket}
\usepackage{bm}
\usepackage{lineno}

\title{Computing Jet Transport Coefficients On The Lattice}

\author[a,b]{Amit Kumar}
\author[b]{Abhijit Majumder}
\author[b]{Ismail Soudi}
\author*[c]{Johannes H. Weber}

\affiliation[a]{Department of Physics, McGill University, \\
Montreal, QC H3A-2T8, Canada}
\affiliation[b]{Department of Physics and Astronomy, Wayne State University, \\
Detroit, MI 48201, USA}
\affiliation[c]{Institut f\"ur Physik, \& IRIS Adlershof \& RTG2575,  Humboldt-Universit\"at zu Berlin, \\
D-12489 Berlin, Germany}

\emailAdd{johannes.weber@physik.hu-berlin.de}

\abstract{
The leading jet transport coefficients $\hat{q}$ or $\hat{e}_{2}$ encode transverse or 
longitudinal momentum broadening of a hard parton traversing a hot medium. 
Understanding their temperature dependence is key to appreciating the 
observed suppression of high-transverse momentum probes at RHIC or LHC 
collision energies. 
We present a first continuum extrapolated result of $\hat{q}$ computed on 
pure SU(3) lattices with non-trivial temperature dependence different from 
the weak-coupling expectation.

We discuss the formalism published in Refs.~\cite{Kumar:2020wvb, Majumder:2012sh} 
and its challenges and status in view of obtaining $\hat{e}_{2}$ or of 
unquenching the calculation. We consider a hard quark subject to a single 
scattering on the plasma. 
The transport coefficients are factorized in terms of matrix elements given 
as integrals of non-local gauge-covariant gluon field-strength field-strength 
correlators. 
After the analytic continuation to the deep-Euclidean region, the hard scale 
permits to recast these as a series of local, gauge-invariant operators. 
The renormalized leading-twist term in this expansion is closely related to 
static quantities, and is computed on pure SU(3) lattices ($N_{\tau}=4,~6,~8,$ 
and $10$) for a wide range of temperatures, ranging from 200MeV < T < 1GeV. 
Our estimate for the unquenched result in $2+1$-flavor QCD has very similar 
features.
}

\FullConference{HardProbes2023\\
 26-31 March 2023 \\
 Aschaffenburg, Germany\\}

\begin{document}
\maketitle

\section{Introduction}

Jets, heavy quarks and quarkonia are the key observables at sufficiently hard 
scales that permit probing the properties of the hot nuclear medium produced 
in heavy-ion collisions at short distance and time scales. 
These hard probes accumulate modifications corresponding to the different 
stages of the hot medium, as the primordial fireball evolves from 
pre-equilibrated early-time dynamics, through a locally equilibrated plasma 
stage, to an extended late-time stage of an inviscid nuclear liquid. 
One usually attempts to capture the mercurial and complex in-medium dynamics 
of such probes in terms of a small number of transport coefficients. 
Once these are determined either from experimental observation or theoretical 
calculation, the kinematics of the hard probes are otherwise treated in a 
simplified, or even classical manner. 
Given the influence of the strongly-coupled late-time stage, it is clear that 
weak-coupling calculation cannot successfully accommodate the underlying 
physics entering these transport coefficients.

For a jet, the leading transport coefficient related to in-medium energy loss 
is the one due to transverse momentum broadening per unit path length, namely, 
$\hat{q} \equiv \tfrac{\braket{k_{\perp}^2}_{T,L}}{L}$, which coincides with 
the second (transverse) moment of the collision kernel $\tfrac{d^4 W(k)}{d^4k}$. 
Another jet transport coefficient that is generally considered as subleading, 
is the one due to longitudinal momentum broadening per unit length, coined 
$\hat{e}_2 \equiv \tfrac{\braket{k_{z}^2}_{T,L}}{L}$ (or $\hat{q}_L$), wherein 
we have assumed a jet traveling almost along the light cone in the negative 
$z$ direction. 
On the level of these transport coefficients, very different models can be 
compared to perturbative or non-perturbative calculations. These coefficients 
may serve as free parameters or input in phenomenological descriptions of 
heavy-ion collision events such as in the JETSCAPE framework~\cite{Soltz:2019aea}. 

\section{Hard parton at leading order}

Both coefficients are obtained from $t$-channel processes, and involve at 
tree-level scattering mediated by one-parton-exchange; 
at leading order in the weak-coupling approach, $\hat{q}$ for a hard parton 
in representation $R$ is a function of the UV and IR cutoffs~\cite{Arnold:2008vd}\footnote{
The constants $\sigma_\pm$ are given in Ref.~\cite{Arnold:2008vd}, 
but are of no importance for the current discussion.}
\begin{equation}
\begin{split}
  \hat{q}(\mu_\text{UV}) &= 
  C_R \sum\limits_{s=\pm}\Xi_s \mathcal{I}_s(\mu_\text{UV}) 
  \frac{g^4 {T^3}}{\pi^2}~, \\
  \mathcal{I}_s(\mu_\text{UV}) &\simeq 
  \frac{\zeta_s(3)}{2\pi} \ln\left(\frac{\mu_\text{UV}}{\mu_\text{IR}}\right) 
  + \Delta \mathcal{I}_s~, \\
  \Delta \mathcal{I}_s &= 
  \frac{\zeta_s(2)-\zeta_s(3)}{2\pi} \left[ \ln\left(\frac{T}{\mu_\text{IR}}\right)
  +\frac{1}{2} - \gamma_E + \ln(2)
  \right] - \frac{\sigma_s}{2\pi}~, \\
  &\Xi_+ = 2N_c,~\Xi_-=2N_f,~\zeta_{\pm}(n) = \sum\limits_{k=1}^{\infty} 
  \frac{(\pm1)^{k-1}}{k^n}~.
\end{split}
\label{eq:qhat finite htl}
\end{equation}
where the IR cutoff is taken to be the Debye mass $\mu_\text{IR} = m_D$, and 
strict weak-coupling hierarchies are assumed: 
$\Lambda_{\text{QCD}} \ll m_D \sim gT \ll T$. 
With each logarithm one associated power of $g^2$ is at the harder and another 
at the softer scale.
Thus, as the UV cutoff $\mu_\text{UV}$ is sent to infinity, i.e. the parton 
is considered to be infinitely hard, the result remains finite and is 
effectively promoted to order $\mathcal{O}(g^2)$, since the 
logarithmic divergence cancels against a running coupling $g^2(\mu_\text{UV})$
at leading order~\cite{Peshier:2008zz, Arnold:2008vd}:
\begin{equation}
  \lim\limits_{\mu_\text{UV} \to \infty} \ln\left(\frac{\mu_\text{UV}}{\mu_\text{IR}}\right) g^2(\mu_\text{UV}) g^2(\mu_\text{IR})
  = \frac{g^2(\mu_\text{IR})}{-2\bar{b}_0}
  \quad\text{with}\quad
  \bar{b}_0 = -\frac{11N_c-2N_f}{N_C (4\pi)^2}~.
\end{equation}
As such the result in the limit of a hard parton is given up to terms of 
order $\mathcal{O}(g^4)$ by 
\begin{equation}
\begin{split}
  \frac{\hat{q}(\infty)}{T^3}
  \equiv
  \lim\limits_{\mu_\text{UV} \to \infty} 
  \frac{\hat{q}(\mu_\text{UV})}{T^3}
&= 
  C_R \sum\limits_{s=\pm} \frac{\Xi_s\zeta_s(3)}{-4\pi^3 \bar{b}_0} 
  g^2(\mu_\text{IR})
  \\
  &=
  C_R \sum\limits_{s=\pm}\frac{\Xi_s\zeta_s(3)}{\Xi_s\zeta_s(2)}~\frac{2N_c \frac{\pi^2}{6}}{-4\pi^3 \bar{b}_0} 
  \left(\frac{m_D(\mu_\text{IR})}{T}\right)^2
  ~.
\end{split}
\end{equation}
For a SU(3) pure gauge plasma, i.e. $N_f=0$, the result becomes particularly 
simple\footnote{We have used the Ramanujan series truncated at the leading 
term for Ap\'ery's constant, 
$\zeta(3) = \frac{7\pi^3}{180} \simeq 1.202\ldots$.}, and is related to the 
$N_f=0$ entropy density in the Stefan-Boltzmann limit,  
$s_\mathrm{SB} = \frac{32}{45}\pi^2{T^3}$, as
\begin{equation}
  \frac{\hat{q}(\infty)}{T^3}
  = 
  \frac{21}{176} ~C_R~ 
  \left(\frac{m_D(\mu_\text{IR})}{T}\right)^2
  \frac{s_\mathrm{SB}}{T^3}
  = 
  \frac{21}{44}\pi ~C_R~ 
  \alpha_s(\mu_\text{IR})
  \frac{s_\mathrm{SB}}{T^3}
  ~.
  \label{eq:qhat infty htl}
\end{equation}
This form is very suggestive: there is a coefficient of order one, a 
representation-dependent Casimir, $C_R$, a factor that accounts for 
medium-modified interactions in any single scattering event, 
$\left(\tfrac{m_D(\mu_\text{IR})}{T}\right)^2$ or 
$\alpha_s(\mu_\text{IR})$, 
and a term from the equation of state (EoS) that accounts for the density 
of available scattering centers, $\tfrac{s_\mathrm{SB}}{T^3}$. 
The counterparts of the latter two are accessible in non-perturbative 
lattice calculations. 
In the physical world, however, i.e. $N_f > 0$, $N_f$ dependent terms break 
the simple relation with the EoS. 
Do similar results arise, if weak coupling---clearly inappropriate for 
media of phenomenological interest at 
$T \simeq gT \simeq \Lambda_{\text{QCD}}$---does not apply?

\section{Formalism}

We use the setup of a hard parton propagating in the negative $z$-direction 
with light-cone momentum $q^{-}$ and follow the formalism outlined in 
Refs.~\cite{Kumar:2020wvb,Majumder:2012sh}.  
At leading virtuality the transport coefficient for $j$-momentum broadening 
is obtained (in $A^{-}=0$ gauge, assuming ergodicity, dropping subleading 
virtualities, promoting $\partial_j \to D_j$, and enforcing an on-shell 
condition) as \vskip-1ex
{\begin{align}
\hat{q}_j(q^{-})
&= 
\sum_{n} \frac{e^{-\beta E_n}}{Z T_{I}} \int {d^4k} k_j^2 
\frac{d^4 W(k)}{d^4k}
\\
&\simeq
c_{0R} g^2
\int \frac{d y^- d^2 y_\perp d^2 k_\perp}{(2\pi)^3} 
e^{i \bm{k}_{\perp} \cdot \bm{y}_{\perp}-i\tfrac{\bm{k}_{\perp}^2}{q^-}y^-}
\Braket{ \mathrm{Tr} \big[ F^{+j}(0) F^{+}_{~j}{(y^-,y_\perp)} \big] }_{T}
~,
\end{align}}
where $c_{0R}=\sqrt{2}C_R$, 
and $g^2$ is the squared gauge coupling at an appropriate thermal IR scale 
$\mu_\text{IR}$. 
Obviously, the full product must be renormalized consistently. 
The thermal correlator $\Braket{ \mathrm{Tr} \big[ F^{+j}(0) F^{+}_{~j}{(y^-,y_\perp)} \big] }_{T}$ is gauge covariant, but at an almost 
light-cone separation. 
The necessary infinitely extended light-cone and transverse Wilson 
lines~\cite{Garcia-Echevarria:2011ewi} imply that a Euclidean definition 
of this quantity---and in particularly one on a hypercubic lattice---is 
far from straightforward. 

For this reason we define a generalized coefficient without enforcing 
the on-shell condition, 
\begin{equation}
\hat{Q}_j(q^+,q^{-}) 
\simeq
c_{0R} g^2
\int \frac{d^4y d^4 k}{(2\pi)^4} 
\frac{e^{i k \cdot y}~2q^-}{(q+k)^2+i\varepsilon}
\Braket{ \mathrm{Tr} \big[ F^{+j}(0) F^{+}_{~j}{(y)} \big] }_{T},
\end{equation}
whose thermal discontinuity at $q^+ \simeq \pm T$ corresponds to 
$\hat{q}_j(q^{-})$. 
For space-like momenta $q^+ \simeq - q^{-}$ there is no nearby d
iscontinuity, and the ratio can be expanded as a geometric series 
in $\tfrac{k_3}{q^{-}}$ (neglecting terms of order $k^2$). 
After promoting $\partial_3 \to D_3$, both integrals can be performed 
and we obtain 
\begin{equation}
\hat{Q}_j(q^+ = -q^{-},q^{-}) 
\simeq
\frac{c_{0R} g^2}{q^{-}} 
\sum\limits_{n=0}^{\infty}
\left(\tfrac{\nu}{q^-}\right)^{n}
\Braket{ \mathrm{Tr} \big[ F^{+j}(0) {\Delta^{n}} F^{+}_{~j}(0) \big] }_{T},
\quad
\Delta \equiv \frac{i\sqrt{2}D_3}{\nu}~,
\end{equation}
where we have introduced a intermediate scale between the thermal IR scale 
and the hard scale, $\mu_\text{IR} \lesssim \nu \ll q^{-}$. 
For a medium that is invariant under parity, odd orders vanish; and for a 
medium at rest there are no mixed terms \`a la $S^3 = F^{0j}F^{3}_{j}$. 

While complex contour integration of 
$\tfrac{\hat{Q}_j(q^+,q^{-})}{q^{-}+q^{+}}$ along a small circle around 
$q^+ \simeq -q^{-}$ yields $\hat{Q}_j(q^+ = -q^{-},q^{-})$, the same 
integral yields for a contour deformed over the whole complex plane the 
contributions from the two branch cuts of $\hat{Q}_j(q^+,q^{-})$, the 
thermal one of width\footnote{The factor $\sqrt{2}$ arises due to 
light-cone coordinates.} ${\delta}_{T}T \simeq 2\sqrt{2}T$ at 
$q^+ \simeq \pm T$, and another one at $q^{+} \ge 0$ for a time-like 
parton undergoing vacuum-like forward splitting. 
The latter vanishes at $n=0$ (at $q^{-} \to \infty$) for a single 
radiated on-shell gluon; beyond that approximation (or beyond $n=0$), 
it can be tamed to some extent through vacuum subtraction, where there 
is no thermal discontinuity. 
Putting everything together we obtain
\begin{equation}
\ensuremath{
\left.\frac{\hat{q}_j(q^{-})}{T^3}
\simeq
c_{0R} g^2 \frac{T}{T_\delta}
\sum_{n=0}^{\infty} \left(\frac{\nu}{q^-}\right)^{2n}
\frac{1}{T^4} \Braket{ \mathrm{Tr} \big[ F^{+j} {\Delta^{2n}} F^{+}_{~j} \big] }_{(T-V)}\right.
}~.
\label{eq:qhatj series}
\end{equation}
The right hand side is a series of vacuum-subtracted, gauge-invariant local 
operators, and thus, in principle, amenable to a lattice calculation. 
$\nu$ could be any scale of the order of the inverse lattice spacing, i.e. 
the temperature $T=\frac{1}{aN_\tau}$. 
Thus, for $q^{-} \to \infty$, the continuum limit at fixed temperature can 
be defined (after appropriate renormalization).

For $q^{-} < \infty$ there are two major problems: first, sending the typical 
scale of the hardest modes, i.e. the cutoff $a^{-1}$, to infinity, implies 
that the geometric series in $\tfrac{k_3}{q^{-}}$ cannot be used. 
Second, the mixing of the temperature-dependent higher-twist operators 
($n > 0$) with temperature-dependent lower dimensional operators multiplied 
by powers of the lattice cutoff cannot be canceled by the vacuum subtraction. 
To date, this unsolved problem is a limitation of our formalism. 

\section{Numerical Results}
\enlargethispage\baselineskip

After a Wick rotation, $x^{0} \!\rightarrow\! -ix^4, A^{0,a} \!\rightarrow\! 
iA^{4,a}\! \Longrightarrow \!F^{0j,a}\! \rightarrow \!iF_{4j}^{a}$, 
which takes\footnote{The factor $\tfrac{1}{4}$ is due to the color trace of 
the generators and the reversal from light-cone to cartesian coordinates.} 
Eq.~\eqref{eq:qhatj series} into 
\begin{equation}
\ensuremath{
\left.\frac{\hat{q}_j(q^{-})}{T^3}
\simeq
\frac{c_{0R}}{4\delta_T}
\sum_{n=0}^{\infty} \left(\frac{T}{q^-}\right)^{2n}
[g^2 \widehat{O}_{j;n}]^{(R)}\right.
~,~ \widehat{O}_{j;n} \equiv \frac{1}{T^4} 
\Braket{ \big[ F_{3j}^{a} {\Delta^{2n}} F_{3j}^{a} -F_{4j}^{a} {\Delta^{2n}} F_{4j}^{a} \big] }_{(T-V)}
}~,
\label{eq:qhatj euclidean}
\end{equation}
we evaluate the leading operator ($\widehat{O}_{j;0}$) on the lattice. 

In pure SU(3) theory, the transverse sum ($j=1,2$) of this operator coincides 
with 
the bare, vacuum-subtracted energy-momentum tensor (EMT) 
$T^{(3)}_{G,34} = \frac{1}{T^4} \sum_\mu
\Braket{ \big[ F_{3\mu}^{a} F_{3\mu}^{a} -F_{4\mu}^{a} F_{4\mu}^{a} \big] }_{(T-V)}
$ 
(in temperature units) in triplet representation\footnote{On a lattice, the continuum EMT's nonet representation 
breaks apart into a sextet and a triplet.}. 
We use plaquette action with $N_\tau=4,~6,~8,$~and~$10$, aspect ratio 
$\tfrac{N_\sigma}{N_\tau}=4$, and the field strength's clover discretization, 
which is a combination whose multiplicative renormalization constant 
$Z^{(3)}_{T}$ is known from finite momentum Ward identities~\cite{Giusti:2015daa, Giusti:2016iqr}. 
In the rest frame,  $T^{(3,R)}_{G,34} = T s$ with the entropy density $s$. 
Thus, we obtain after a few approximations 
\begin{equation}
\ensuremath{
\left.\frac{\hat{q}(\infty)}{T^3}
\simeq
\frac{c_{0R}}{4\delta_T} [g^2]^{(R)} \frac{s}{T^3} 
=
\frac{1}{8}
~C_R~ 
[g^2]^{(R)} \frac{s}{T^3} 
\right.
}~,
\label{eq:qhat euclidean}
\end{equation}
which is extremely similar to the LO result of Eq.~\eqref{eq:qhat infty htl}. 
Since the entropy density is a (scheme-independent) physical observable, 
the scheme choice for the squared coupling introduces a scheme 
dependence for $\hat{q}(\infty)$. 
We use $N_f=0$ $\overline{\textrm{MS}}$ one-loop coupling at 
the scale $\mu_{\text{IR}} = (2\ldots4)\pi T$.
\enlargethispage\baselineskip

\begin{figure}[t]
\centering
\includegraphics[width=0.5\textwidth]{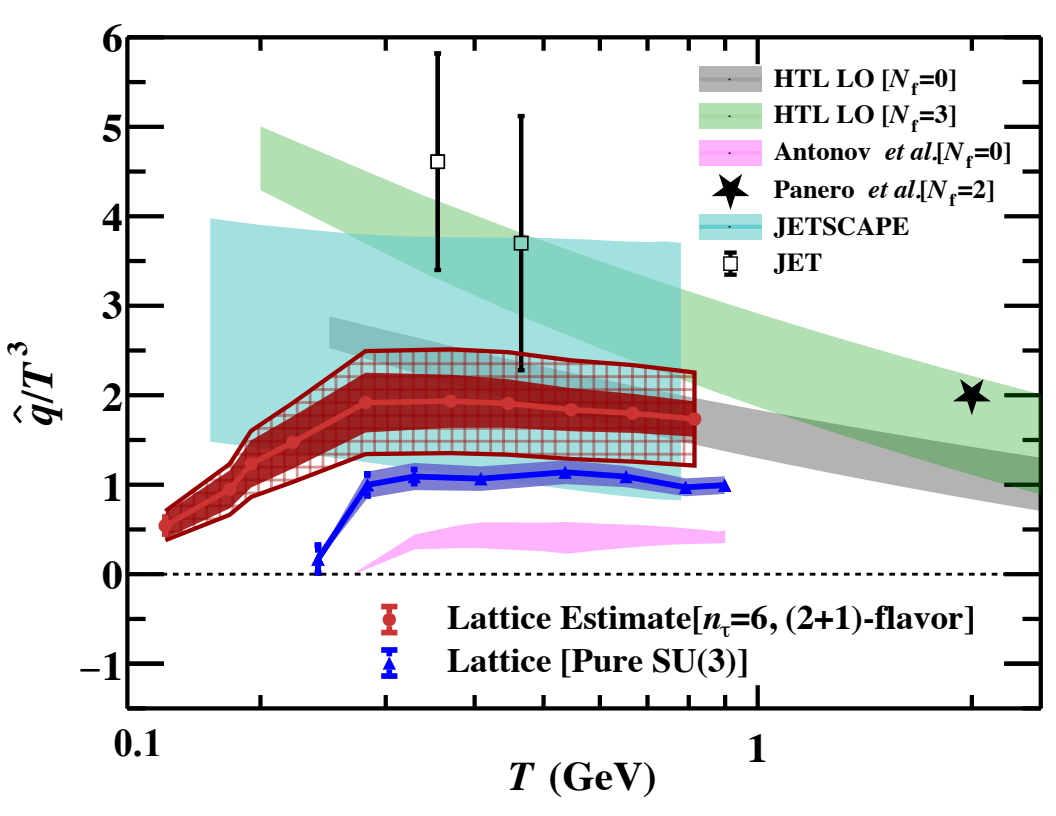}%
\caption{\label{fig:qhat}%
$\tfrac{\hat{q}(\infty)}{T^3}$ calculated on a lattice rises similarly to 
$\tfrac{s}{T^3}$ during the QCD crossover/transition, and exhibits a rather 
flat temperature dependence above $T \gtrsim 300$ MeV. 
Phenomenological determinations~\cite{JET:2013cls, Soltz:2019aea} are 
quantitatively close,  
and a stochastic vacuum model finds a similar trend~\cite{Antonov:2007sh}. 
LO HTL at $q^- = 100$ GeV assuming constant $g^2(\mu_{\text{IR}})$, i.e. 
the $\mathcal{O}(g^4)$ contribution in Eq.~\eqref{eq:qhat finite htl}, 
rises logarithmically towards lower temperatures and is compatible only at 
$T \gtrsim 2$ GeV, where the NNLO EQCD result~\cite{Panero:2013pla} is 
similar, too. 
}
\end{figure}

In full QCD, the identification with the EMT fails, since $[\sum_j\widehat{O}_{j;0}]^{(R)}$ corresponds only to the renormalized 
EMT's gauge part. 
In light of the LO results, lack of a simple relation to the EoS is 
unsurprising. 
Bare gauge and quark parts mix upon renormalization, where not all mixing 
matrix elements could be fixed from Ward identities. 
They depend on the details of gauge and quark actions, and are 
unknown\footnote{It is known at one-loop order for some other actions, 
see~\cite{DallaBrida:2020gux}. We use this to estimate the size of missing 
contributions.} for the combination of actions we use, namely, 
L\"uscher-Weisz gauge action and (2+1)-flavors of highly improved 
staggered quarks (HISQ) (physical strange quark, and $m_\pi=161$ MeV in 
the continuum limit). We use $N_\tau=4,~6,$~and~$8$, aspect ratio 
$\tfrac{N_\sigma}{N_\tau}=4$, and again the clover discretization. 
For approximate renormalization we apply tadpole 
improvement~\cite{Banerjee:2011ra} of the QCD results, and estimate the 
missing contributions are with two complementary arguments. 
Overall, this leads to a 30\% systematic uncertainty; see 
Ref.~\cite{Kumar:2020wvb} for details.
We use $N_f=3$ $\overline{\textrm{MS}}$ one-loop coupling at the scale 
$\mu_{\text{IR}} = (2\ldots4)\pi T$, and show our estimate (at $N_\tau=6$) 
together with the $N_f=0$ continuum result in Fig.~\ref{fig:qhat}.

Applying the same formalism, we may obtain $\hat{e}_{2}$ by setting $j=3$. 
The first (magnetic) term of $\widehat{O}_{3;0}$ vanishes exactly, and 
we have to recombine the second (electric) one to produce two 
scale-independent contributions in pure SU(3) theory. 
As a consequence, $\hat{e}_{2}$ also receives a contribution from the 
singlet representation of the vacuum-subtracted EMT that does not share the 
scheme dependence of $\hat{q}(\infty)$, namely,   
\begin{equation}
\ensuremath{
\left.\frac{\hat{e}_2(\infty)}{T^3}
\simeq
\frac{1}{4}\frac{\hat{q}(\infty)}{T^3}
+\frac{2\pi^2}{3N_C b_0}\frac{T^{(1)}_G}{T^4}
\right.
}~,
\label{eq:e2hat euclidean}
\end{equation}
which is also accessible through lattice calculations, and subject of ongoing work. 

\section{Outlook and discussion}

We have demonstrated that both $\hat{q}$ and $\hat{e}_2$ are in principle 
amenable to a lattice calculation for $q^{-} \to \infty$, and can be related 
to the EoS in pure SU(3) theory. 
The formal similarity to LO results, Eq.~\eqref{eq:qhat infty htl} 
and the closeness to phenomenology are astonishing. 
The logarithmic rise of HTL appears to be suppressed in the lattice result due 
to a diminishing number of scatterers at $T \lesssim 300$ MeV. 

\vskip1ex
\textbf{Acknowledgments:} 
This work was supported by the National Science Foundation under grant No. ACI-1550300 
within the JETSCAPE collaboration and by the Department of Energy under grant No. DE-SC0013460.
J.H.W.’s research is funded by the Deutsche Forschungsgemeinschaft (DFG, German Research
Foundation)---Projektnummer 417533893/GRK2575 ``Rethinking Quantum Field Theory''.

\end{document}